\documentclass{article}

\PassOptionsToPackage{numbers, compress}{natbib}

\usepackage[preprint]{neurips_2022}

\usepackage[utf8]{inputenc} 
\usepackage[T1]{fontenc}    
\usepackage{hyperref}       
\usepackage{url}            
\usepackage{booktabs}       
\usepackage{amsfonts}       
\usepackage{nicefrac}       
\usepackage{microtype}      
\usepackage{xcolor}         
\usepackage{amsmath}
\usepackage{verbatim}

\usepackage[pdftex]{graphicx}
\usepackage{bm}

\renewcommand{\vec}[1]{\bm{\mathbf{#1}}}
\newcommand{\mat}[1]{\bm{\mathbf{#1}}}
\newcommand{\norm}[1]{\left\lVert#1\right\rVert}

\title{Action-modulated midbrain dopamine activity arises from distributed control policies}

\author{%
  Jack Lindsey \\
  Department of Neuroscience\\
  Columbia University\\
  New York, NY \\
  \texttt{jackwlindsey@gmail.com} \\
   \And
   Ashok Litwin-Kumar \\
  Department of Neuroscience\\
  Columbia University\\
  New York, NY \\
  \texttt{a.litwin-kumar@columbia.edu} \\
}

\begin{document}

\maketitle

\begin{abstract}
Animal behavior is driven by multiple brain regions working in parallel with distinct control policies. We present a biologically plausible model of off-policy reinforcement learning in the basal ganglia, which enables learning in such an architecture. The model accounts for action-related modulation of dopamine activity that is not captured by previous models that implement on-policy algorithms.  In particular, the model predicts that dopamine activity signals a combination of reward prediction error (as in classic models) and ``action surprise," a measure of how unexpected an action is relative to the basal ganglia's current policy. In the presence of the action surprise term, the model implements an approximate form of $Q$-learning.  On benchmark navigation and reaching tasks, we show empirically that this model is capable of learning from data driven completely or in part by other policies (e.g. from other brain regions).  By contrast, models without the action surprise term suffer in the presence of additional policies, and are incapable of learning at all from behavior that is completely externally driven.  The model provides a computational account for numerous experimental findings about dopamine activity that cannot be explained by classic models of reinforcement learning in the basal ganglia.  These include differing levels of action surprise signals in dorsal and ventral striatum, decreasing amounts movement-modulated dopamine activity with practice, and representations of action initiation and kinematics in dopamine activity. It also provides further predictions that can be tested with recordings of striatal dopamine activity. \footnote{Source code for our experiments to be provided upon publication.}

\end{abstract}

\section{Introduction}

An extensive body of work has sought to account for the function of the basal ganglia using the computational framework of reinforcement learning (RL), in particular RL algorithms for action selection and value learning \citep{joel2002actor, niv2009reinforcement, ito2011multiple, sutton2018reinforcement}.  
Striatal neurons within the basal ganglia integrate diverse inputs, including projections from across the cerebral cortex.
The activity of these neurons, in particular neurons in the dorsal striatum, plays a key role in action selection \citep{packard2002learning}.  
On the other hand, neurons in the ventral striatum have been shown to encode the learned value of stimuli \citep{cardinal2002emotion, montague1996framework, o2004dissociable}.  
The phasic activity of midbrain dopamine neurons projecting to the striatum gates plasticity at cortico-striatal synapses \citep{wickens1996dopamine, calabresi2000dopamine, contreras1999predictive} and may therefore modulate the learning of actions and values.
Indeed, in many RL algorithms, the learning of state-to-action mappings (``policies'') and value estimates is modulated by a scalar factor known as the ``advantage'' or ``reward prediction error'' (RPE), which measures deviations in attained reward from expectations based on learned value estimates. 
Numerous experiments have shown that striatal dopamine activity encodes an RPE-like signal \citep{schultz1997neural, montague1996framework, houk199513}.  

Collectively, these findings suggest a model in which the basal ganglia implements an online actor-critic RL algorithm, where ventral and dorsal striatal subregions play the roles of critic and actor, respectively, and dopaminergic activity encodes the advantage (RPE) signal \citep{joel2002actor}.  This model has been extended in a variety of ways to incorporate more biological detail \citep{suri1998learning, suri1999neural, contreras1999predictive, brown1999basal, suri2001modeling}, ideas from model-based RL \citep{daw2011model}, distributional RL \citep{dabney2020distributional}, and meta-RL \citep{wang2018prefrontal}.

Despite these promising links, there remain challenges to the view of basal ganglia as implementing an actor-critic RL algorithm.  In this work, we address two such challenges. First, dopamine activity in the basal ganglia is observed to encode other information beside RPEs \citep{engelhard2019specialized}, particularly signals relating to movement initiation and vigor \citep{da2018dopamine, zenon2016dopamine}.  

Second, the classic actor-critic model of the basal ganglia is an on-policy RL algorithm---it is designed to learn online from experiences driven by the basal ganglia's policy.  However, motor control in biological systems is distributed, with multiple brain regions including the motor cortex and cerebellum exerting influence on behavior independent of the basal ganglia \citep{exner2002differential, wildgruber2001differential, ashby2010cortical, silveri2021contribution, bostan2018basal}. We show that classic actor-critic models of the basal ganglia fail in this scenario and argue that the ability to learn from off-policy experience is essential to models of basal ganglia learning.

We present a biologically plausible model of off-policy RL in continuous action spaces.  Our model differs from the classic actor-critic model of the basal ganglia by adding an additional action-related term to dopamine activity.  We call this term ``action surprise," as it measures deviation of the action an agent takes in a particular state from the typical action that the output of the basal ganglia would select in that state.  We show mathematically that, with this addition, the algorithm implements an approximate form of $Q$-learning in continuous action spaces proposed by \citep{gu2016continuous}.  Using simulations, we show that action surprise is essential to effective learning when controllers other than the basal ganglia also contribute to behavior. Thus, action-related activity need not be understood as an independent function of midbrain dopamine neurons separate from their role in learning, but rather as a necessary component of an algorithm that enables off-policy RL. This action surprise model accounts for several experimental findings about movement-related dopamine activity in the basal ganglia. We also provide an interpretation of the presence of greater levels of movement-related activity in dorsal-projecting dopamine neurons, showing that such an asymmetry can approximate a supervised learning signal that accelerates the consolidation of other controllers' policies. Finally, we describe predictions of our model regarding dopamine activity.

\section{Background: Connection between policy gradient algorithms and cortico-striatal plasticity}
\label{section:background}
Here we briefly introduce the connection between on-policy policy gradient algorithms and synaptic plasticity rules in the striatum. In what follows, we assume familiarity with standard notation and concepts in RL; for a brief review, see Appendix \ref{appendix:notation}. Policy gradient algorithms update the parameters $\vec{\theta}_P$ of the policy according to
\begin{equation}
\label{eq:policygradient}
   \Delta \vec{\theta}_P \propto \nabla_{\vec{\theta}_P} \log \pi (\vec{a}_t|\vec{s}_t) \delta_t,
\end{equation}
where $\delta_t$ is an estimate of the advantage function $A(\vec{s}_t, \vec{a}_t) = Q(\vec{s}_t, \vec{a}_t) - V(\vec{s}_t)$, which will be discussed in more detail below.

Throughout this work we assume that policies $\pi(\vec{a}|\vec{s})$ are parameterized by Gaussian distributions $a_i \sim \mathcal{N}(\mu_i(\vec{s}), \sigma_i^2)$. For simplicity, in what follows we assume  $\sigma_i = \sigma$ for all $i$; however, our results are easily extended to the case of $\sigma_i$ varying across action dimensions. Letting $\vec{\theta}_{\vec{\mu}}$ be the parameters of $\vec{\mu}(\vec{s})$,  we have

\begin{equation}
\label{eq:on_policy_actor_update}
\Delta \vec{\theta}_{\vec{\mu}} \propto  \frac{1}{\sigma^2} \delta_t \left (\vec{a}_t - \vec{\mu}(\vec{s}_t) \right ) \nabla_{\vec{\theta}_{\vec{\mu}}} \vec{\mu}(\vec{s}_t).
\end{equation}
Supposing that our policy is parameterized by a linear map from some feature representation of $\vec{s}_t$, $\vec{\mu}(\vec{s}_t) = \mat{W}_{\vec{\mu}} \vec{\phi}(\vec{s}_t)$, this becomes:
\begin{equation}
\label{eq:threefactor}
\Delta \mat{W}_{\vec{\mu}} \propto  \frac{1}{\sigma^2} \delta_t \left (\vec{a}_t - \vec{\mu}(\vec{s}_t) \right ) \vec{\phi}(\vec{s}_t)^\top.
\end{equation}
This is a ``three-factor'' learning rule \citep{fremaux2016neuromodulated} for synapse $W_{ij}$ obtained by multiplying presynaptic activations $\phi_j(\vec{s}_t)$, a postsynaptic term $(\vec{a}_t)_i - \mu_i(\vec{s}_t)$ measuring the deviation of the sampled action from the typical action in this state, and a third factor $\delta_t$. RL models of the basal ganglia assume that $(\vec{a}_t)_i - \mu_i(\vec{s}_t)$ is available to the postsynaptic neuron and that $\delta_t$ is signaled by dopamine release in the striatum \citep{sutton2018reinforcement, niv2009reinforcement}. Consistent with this model, experimental observations have shown that a coincidence of dopamine release and pre and post-synaptic neural activity is necessary for plasticity at cortico-striatal synapses \citep{wickens1996dopamine, calabresi2000dopamine, contreras1999predictive}. Specific biological implementations of this learning rule are discussed further in Appendix \ref{appendix:threefactor}.

In actor-critic models the $\delta_t$ factor is
\begin{equation}
\delta_t = r_{t+1} + \gamma \hat{V}(\vec{s}_{t+1}) - \hat{V}(\vec{s}_t) ,
\end{equation}
often referred to as ``reward prediction error'' (RPE). Here, $\hat{V}$ is an estimate of the value function output by a ``critic'' network, separate from the policy network, which learns its parameters $\vec{\theta}_V$ using temporal difference (TD) learning:
\begin{equation}
\label{eq:on_policy_critic_update}
\Delta \vec{\theta}_V \propto \delta_t \nabla_{\vec{\theta}_V} \hat{V}(\vec{s}_t).
\end{equation}

In models of the basal ganglia, the ventral striatum is often assigned the role of the critic, as it is implicated in value-learning but less so in controlling actions \citep{montague1996framework, niv2009reinforcement}. Importantly, TD learning uses the same quantity $\delta_t$ for learning the value function as for policy learning.  Hence, a scalar $\delta_t$ signal measuring RPE and broadcast across the striatum supports learning for both the actor and critic. RPE captures key experimental features of striatal dopamine activity---responses to unexpected reward or reward-predictive cues, no significant response to expected reward, and suppression in response to unexpected lack of reward \citep{schultz1997neural, montague1996framework, houk199513}.

\section{Off-policy RL through action surprise signals in dopamine activity}

The actor-critic algorithm described above is an on-policy algorithm.  
Hence, the algorithm may perform suboptimally, or fail altogether, if the actions used for learning are not (or not always) sampled from the learned policy. Off-policy algorithms are often used to enable learning from a replay buffer of past experiences, expert demonstrations, and/or a separate exploration policy \citep{arulkumaran2017deep}. In a biological context, we argue that the fact that other brain regions can exert control over behavior independently of the basal ganglia further motivates off-policy RL. Indeed, later we show empirically that standard on-policy algorithms suffer when additional controllers exert (partial) control over an agent's behavior.  

\subsection{Action-sensitive dopamine activity arises from parameterized $Q$-learning}
 
While there are a variety of approaches to off-policy RL, most require learning an estimate of the $Q$-function $Q(\vec{s}, \vec{a})$ rather than the state-value function $V(\vec{s})$.  $Q$-learning iteratively minimizes the following loss function:
\begin{align}
\label{eq:Qlearning}
\mathcal{L} &= \norm{y_{t+1}  - \hat{Q}(\vec{s}_t, \vec{a}_t)}^2, \\
y_{t+1} &=  r_{t+1} + \gamma \max_{\vec{a}} \hat{Q}(\vec{s}_{t+1}, \vec{a}).
\end{align}
Computing the quantity $\max_{\vec{a}} Q(\vec{s}, \vec{a})$ directly is intractable in high-dimensional, continuous action spaces. A variety of approaches to this problem have been proposed. Here we focus on an approach adopted by \cite{gu2016continuous}, which involves restricting the form of the $Q$-function estimate to a family of functions whose maximum is easy to compute (we discuss alternatives which have proven to be effective in deep RL, but are difficult to map to biological implementations, in Appendix \ref{appendix:offpolicy}). 

We parameterize the $Q$-function as follows:
\begin{equation}\
\label{eq:parameterization}
\hat{Q}(\vec{s}, \vec{a}) = \hat{V}(\vec{s}) - \frac{1}{\sigma^2}\norm{\vec{a} - \vec{\mu}(\vec{s})}^2.
\end{equation}
For now we treat the the scaling factor $\sigma$ as a fixed hyperparameter; however, it can also be learned online if desired (see Appendix \ref{appendix:confidence}). This parameterization can be considered  a special case of the parameterization of \cite{gu2016continuous}.

A primary insight of our work is the observation that, under the parameterization of Eq.~\ref{eq:parameterization}, gradient updates of the loss function of Eq.~\ref{eq:Qlearning} 
yield a biologically plausible actor-critic algorithm with action-sensitive dopamine activity.  In particular, taking the gradient of the $Q$-learning loss function with respect to the parameters $\vec{\theta}_V$ and $\vec{\theta}_{\vec{\mu}}$ of $\hat{V}$ and $\vec{\mu}$, respectively, yields the following learning updates:
\begin{align}
\label{eq:action_surprise_actor_update}
\Delta \vec{\theta}_V &\propto \delta^+_t \nabla_{\vec{\theta}_V} \hat{V}(\vec{s}_t),\\
\Delta \vec{\theta}_{\vec{\mu}} &\propto \frac{1}{\sigma^2} \delta^+_t (\vec{a}_t - \vec{\mu}(\vec{s}_t)) \nabla_{\vec{\theta}_{\vec{\mu}}} \vec{\mu}(\vec{s}_t),
\end{align}
where
\begin{align}
\delta^+_t &= r_{t+1} + \gamma \hat{V}(s_{t+1}) - \hat{V}(\vec{s}_t) + \frac{1}{\sigma^2} \norm{\vec{a}_t - \vec{\mu}(\vec{s}_t)}^2\\
&= \delta_t + \frac{1}{\sigma^2}\norm{\vec{a}_t - \vec{\mu}(\vec{s}_t)}^2.
\end{align}
These update equations are the same as those of the standard on-policy advantage actor-critic algorithm, but with one additional term added to the dopaminergic signal. Now the dopamine factor $\delta^+$ represents a the sum of the classic RPE $\delta$ and $\norm{\vec{a} - \vec{\mu}(\vec{s})}^2$, which measures the deviation of the sampled action from the action that would be most likely to be chosen by the actor. We refer to this term as ``action surprise.'' 

\begin{figure}[h!]
    \centering
    \includegraphics{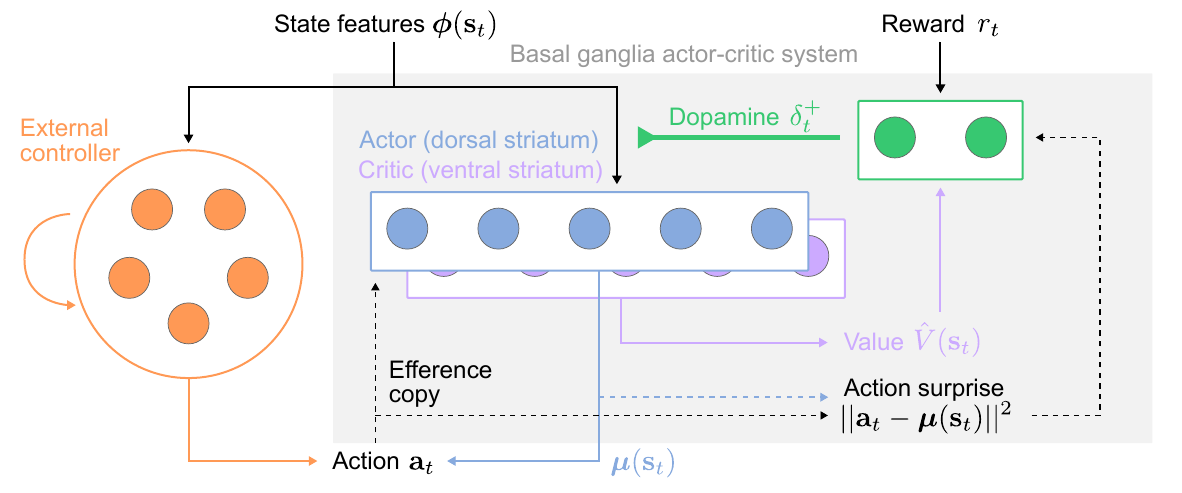}
    \caption{Schematic of model architecture.  Actions are driven in parallel by an actor-critic architecture in the basal ganglia (gray region) and an external controller, representing other brain regions (orange). Both systems receive state information via a feature representation $\vec{\phi}(\vec{s}_t)$ (e.g. cortico-striatal projection neuron activity).  The actor-critic module performs RL with weight updates modulated by $\delta_t^+$, which is a combination of both RPE and action surprise. Dashed lines indicate architectural features of the action surprise model that are not present in the classic actor-critic model.}
    \label{fig:model}
\end{figure}

We note that our model is agnostic to whether action surprise is encoded by the same neurons as RPE.  It may be encoded by separate neurons as long as they release dopamine in the same areas as RPE-signaling dopamine. 
We also note that the update equation (Eq.~\ref{eq:action_surprise_actor_update}) for the actor uses the action $\vec{a}_t$ taken by the agent.  Biologically, this requires an efferent copy of the agent's action (taking into account the influence of other controllers) be sent to the striatal projection neurons representing action in the basal ganglia.  This architecture is consistent with the presence of pathways from motor cortex, thalamus, and cerebellum to striatal projection neurons \citep{melzer2017distinct, haber2022corticostriatal, bostan2010basal, mcfarland2000convergent}.   A schematic of the connections involved in our model is depicted in Fig.~\ref{fig:model}.

\subsection{Differential action surprise signals in dorsal and ventral striatum}

The formulation above predicts that the same dopamine signal $\delta^+$ is broadcast to both the actor and critic.  However, the ratio of movement-related to reward-related modulation of dopamine activity appears to vary across the striatum, with more movement-related activity in dorsal regions \citep{howe2016rapid}. In our modeling framework, this corresponds to the action surprise term being weighted more strongly in the actor (dorsal striatum) than the critic (ventral striatum).   Such an asymmetry is produced by adding an additional term to the weight updates of the actor, proportional to
\begin{equation}
\label{eq:supervision}
  \norm{\vec{a}_t - \vec{\mu}(\vec{s}_t)}^2 (\vec{a}_t - \vec{\mu}(\vec{s}_t)) \nabla_{\vec{\theta}_{\vec{\mu}}} \vec{\mu}(\vec{s}_t).
\end{equation}
This update is aligned with the gradient with respect to $\vec{\theta}_{\vec{\mu}}$ of the following loss function:
\begin{equation}
\mathcal{L}_{\mathrm{sup}} = \mathbb{E}\left [\norm{\vec{a}_t - \vec{\mu}(\vec{s}_t)}^2 \right],
\end{equation}
with the learning rate itself governed by the magnitude of action surprise. Hence, these updates encourage the basal ganglia policy to imitate the agent's behavioral policy on steps with large action surprise.  Heuristically, steps with large action surprise are more likely to have been driven by an external controller.  An additional action surprise contribution to actor-modulating dopamine activity can therefore be interpreted as an approximate supervised learning signal encouraging imitation of the external controller's policy.  We show in simulations that this effect helps the basal ganglia more rapidly consolidate expert policies of other controllers.

\section{Experimental Setup}

\emph{\textbf{Tasks}}. We simulated two simple continuous control tasks to demonstrate the role of the action surprise term in off-policy learning.  See Fig.~\ref{fig:results}A for illustrations.

\emph{Open-field navigation:} An agent is positioned in a two-dimensional square environment with continuous coordinates.  On each trial, the initial position of the agent and a goal location are randomly sampled.  The agent controls its $x$ and $y$ acceleration. 

\emph{Two-joint:} An agent controls a two-joint arm with two equal-length segments.  The arm position and a target location are randomly sampled at the beginning of each trial.  The agent outputs torques at each of the joints in order to move the most peripheral point of the arm toward the target.

In both tasks the cost incurred by the agent at each time step is proportional to the inverse of its squared distance from the goal location $\vec{g}$ at each time step, plus penalties for its squared  velocity and acceleration (summed across joints in the two-joint arm case) \begin{align}r_t =
-\left(\norm{\vec{x}(t) - \vec{g}}^2 + \norm{\alpha_1\frac{d\vec{x}(t)}{dt}}^2 + \norm{\alpha_2\frac{d^2\vec{x}(t)}{dt^2}}^2 \right)\end{align}

\emph{\textbf{Model architecture}}. For both tasks we used a shallow neural network architecture with a single hidden layer.  The state inputs to the network consist of the agent's (angular) position and (angular) velocity, as well as the position of the target location, a total of six scalar variables.  Each of these variables is represented as a one-hot vector by discretizing its domain into 10 equally spaced bins.  These vectors are concatenated and the result used as the network input.  The hidden layer of the network has size $256$ and uses the ReLU nonlinearity, and the input weights to the hidden layer are fixed at their random (Kaiming uniform) initialization.  We fixed these weights to avoid the complexity of biological implementations of backpropagation, since this is not the focus of our work (however, we note that the action surprise model can be applied to deep networks by backpropagating gradients through additional layers).  Thus, the hidden layer activations of the network serve as a fixed feature representation $\vec{\phi}(\vec{s})$ of the environment state, and all learning occurs in output weights $\mat{W}_{\vec{\mu}}$ and $\vec{w}_V$ which output actions $\vec{\mu} = \mat{W}_{\vec{\mu}} \vec{\phi}(\vec{s})$ and value estimates $\hat{V} = \vec{w}_V \cdot \vec{\phi}(\vec{s})$, respectively.  We refer to this network as the basal ganglia network.

\emph{\textbf{External controllers and training protocol}}. To model the influence of other brain regions on behavior, we introduced an additional neural network that also exerted control over the agent's actions (Fig.~\ref{fig:model}, orange). To vary the performance of this controller's policy, we trained the controller on the task with backpropagation for different numbers of steps: 0 (random policy), 2,000 (intermediate policy), or 100,000 (expert policy).  We also varied the degree to which the external controller influences behavior.  In the fully on-policy case, the external controller is not used.  In the fully off-policy case, the external controller entirely drives the behavior.  In the case of partial control, each action is sampled from the basal ganglia's network or the external controller with probability 0.5 each. Additionally, we implemented an alternative partial control mechanism, in which the average of the basal ganglia network output and the external controller output is used.  

Throughout training, Gaussian exploration noise is added to the  output of both networks at each step.  In all cases, for the action surprise model, the action $\vec{a}_t$ used for the actor update in Eq.~\ref{eq:action_surprise_actor_update} is the action taken by the agent, taking into account the contributions of both controllers and the exploration noise.

\emph{\textbf{On-policy baseline models}}. We compared the action surprise model with two baselines in which dopamine neurons signal pure RPE.  The first is described by Eqs. \ref{eq:on_policy_actor_update} and \ref{eq:on_policy_critic_update}, where the action used the actor update is simply the action $\vec{a}_t$ taken by the agent, as in the action surprise model.   As discussed above, this approach is potentially unstable in the presence of other controllers due to the mismatch between the behavioral policy from which actions are sampled and the basal ganglia network's policy.  As a second, potentially more competitive baseline, we instead used the output of the basal ganglia network alone, including exploration noise, in place of $\vec{a}_t$ in Eq.~\ref{eq:on_policy_actor_update}.  We refer to this as the ``no efferent copy'' model since the basal ganglia is blind to the action ultimately taken by the agent.  In this version, the influence of the external controller is effectively treated as part of the environment. We note that this approach has no ability even in principle to learn from fully off-policy data, but can learn when the basal ganglia network exerts partial control over actions.

\emph{\textbf{Hyperparameters}}. For all models we optimized the learning rate and magnitude of exploration noise as hyperparameters.  For the action surprise model we treated the coefficient $\frac{1}{\sigma^2}$ of the action surprise term as a hyperparameter. Since the choice of $\sigma$ affects the overall magnitude of updates in the actor, to avoid biasing results we allowed the actor and critic learning rates to be optimized as separate hyperparameters for the RPE-only models.  See Appendix \ref{appendix:parameters} for details.

\section{Experimental Results}

\begin{figure}[h!]
    \centering
    \includegraphics{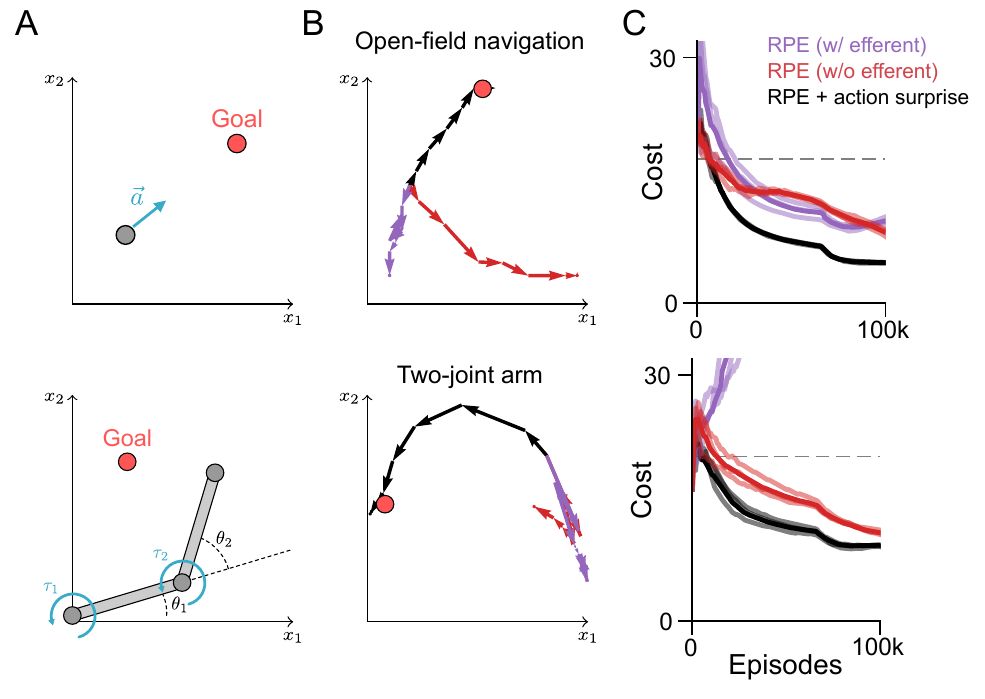}
    \caption{\textbf{A}:~Depictions of simulated tasks. \textbf{B}:~Example trials in  which the action surprise model successfully reaches the target but the RPE-only models fail to.  \textbf{C}:~Performance of the three models when the basal ganglia shares control of behavior with an external controller. Dark lines indicate mean performance over three runs, and faint individual lines (in this case difficult to see due to tight overlap) indicate individual runs.  Performance shown reflects policy of the basal ganglia network and external controller. Dashed line indicates the performance of the external controller alone.}
    \label{fig:results}
\end{figure}

\begin{figure}[h!]
    \centering
    \includegraphics[width=1.0\linewidth]{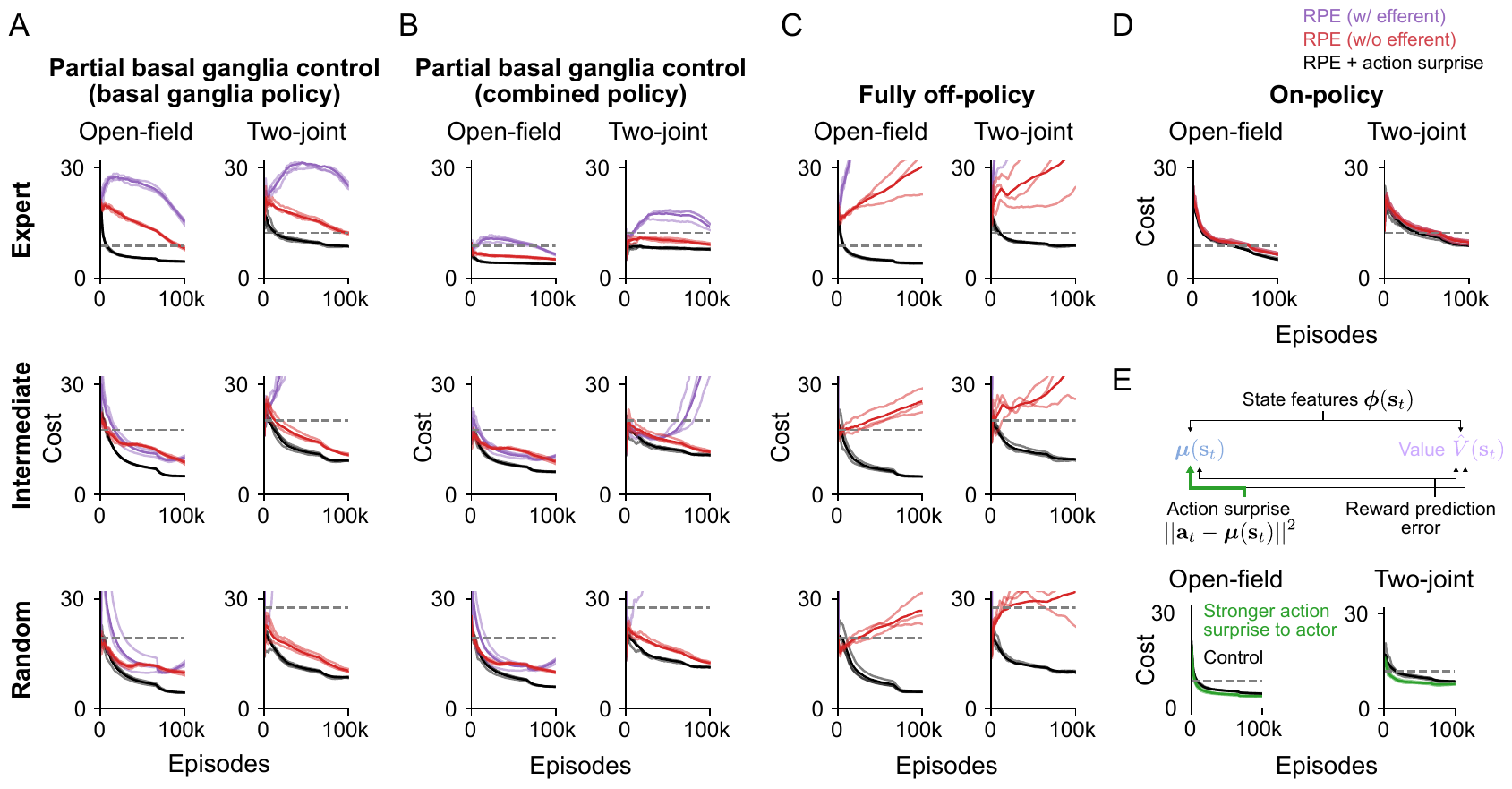}
    \caption{\textbf{A}:~Performance of the three models, as in Fig.~\ref{fig:results}C, shown for external controllers with varying levels of expertise (random, intermediate, expert; trained via backpropagation on the task for different numbers of steps). Middle row corresponds to Fig.~\ref{fig:results}C. \textbf{B}:~Same as A, but showing the performance of the combined policy of the basal ganglia network and external controller.  \textbf{C}:~Same as A, but in the case where all behavior is driven by the external controller during learning.  \textbf{D}:~Performance of the RPE and RPE+action surprise models for fully on-policy learning on the two tasks.  \textbf{E}:~Top: Schematic of model in which the contribution of the action surprise term to dopamine activity is weighted more strongly in the actor than in the critic (green arrow).  Bottom: comparison of this model to the original action surprise model in the case of partial basal ganglia control and fully off-policy learning with the expert controller (black traces are the same as in the top row of panel E). In this example the action surprise term is weighted 8 times as strongly in the actor than the critic.}
    \label{fig:results2}
\end{figure}

We first analyzed performance for the case of actions partially driven by the basal ganglia network and partially driven by the external controller, intended as a representative model of behavioral control distributed across brain regions.   Fig.~\ref{fig:results}C shows that the basal ganglia policy attained superior learning speed and performance than RPE-only models (shown here is a representative case using the intermediate-level external controller). Fig. ~\ref{fig:results}B shows example trials in which this performance advantage manifests very clearly, with the action surprise model successfully reaching the target while the other models fail completely. 
The action surprise model remained advantageous when we varied the level of expertise of the external controller (Fig.~\ref{fig:results2}A).   The same was true when we assessed the combined performance of the basal ganglia network and the external controller (Fig.~\ref{fig:results2}B), rather than the performance of the basal ganglia network by itself.  Notably, the RPE-only model with efferent copy failed entirely to learn in some cases.   We found similar results when the contributions of the basal ganglia network and external controller were averaged rather than being combined by sampling (Appendix \ref{appendix:averagingresults}).

We next examined the fully off-policy case.   The RPE-only model with no efferent copy failed to learn from off-policy data in all cases, as expected, as the action term used in the update for this model is uncorrelated with the actions taken by the external controller.  Interestingly, the RPE-only model with efferent copy also failed to learn in all cases, in fact even more catastrophically.   The action surprise model, by contrast, successfully learned both tasks regardless of the quality of the external controller's policy (Fig.~\ref{fig:results2}C).

A potential concern is that the action surprise model's improvement for off-policy learning comes at the expense of on-policy learning performance.  However, we found that even in the case of fully on-policy learning, with no external controller, the action surprise model matched and indeed slightly outperformed the RPE-only model (Fig.~\ref{fig:results2}D). Note that there is no distinction between the efferent copy/no efferent copy variants in the on-policy case.

We also tested the variant of the action surprise model in which an additional update (Eq.~\ref{eq:supervision}) is applied to the actor, corresponding to a higher action surprise coefficient in the actor (dorsal striatum) than in the critic (ventral striatum). We found that this addition improves learning in the presence of expert external controllers (Fig.~\ref{fig:results2}E), consistent with accelerated consolidation of the expert policy.

\section{Biological implications of the action surprise model}
\label{sec:bio}

\begin{figure}[h!]
    \centering
    \includegraphics[width=\linewidth]{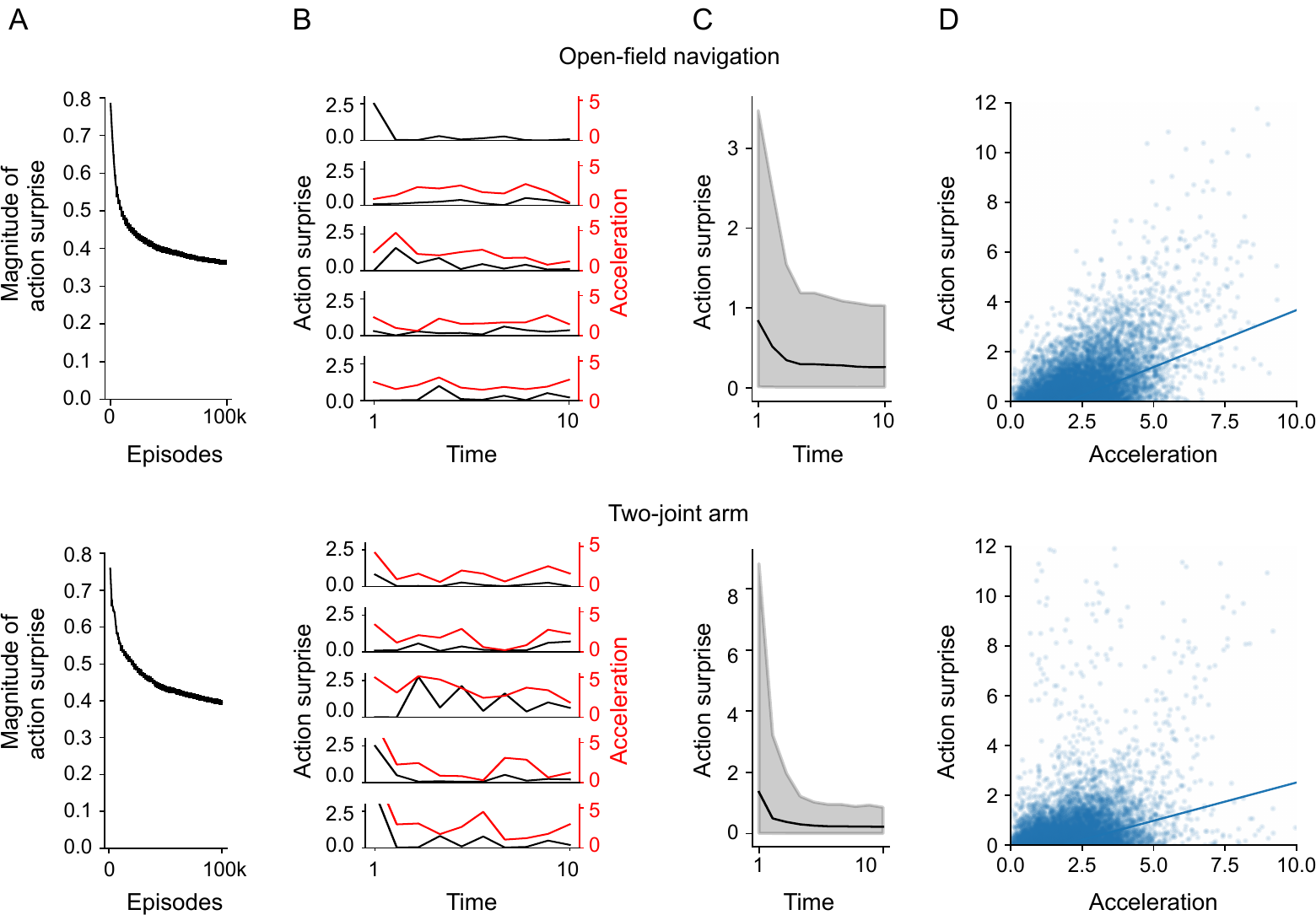}
    \caption{\textbf{A}:~Mean and standard deviation of action surprise term in the dopamine response over the course of learning, in the case where the basal ganglia shares partial control with an expert controller.  In all these examples $\frac{1}{\sigma^2}$ was set to 0.125 and and the exploration noise to $1.0$. \textbf{B}:~Traces of the action surprise signal (black) and the the agent's acceleration (red) -- or rotational acceleration averaged across joints in the two-joint arm case -- for five randomly sampled example trials. \textbf{C}:~Mean and 95\% confidence interval (across trials) of action surprise as a function of the time within a trial.  \textbf{D}:~Scatter plots (showing individual steps, across many episodes) of the action surprise term in the model dopamine response vs. magnitude of agent acceleration. }
    \label{fig:predictions}
\end{figure}

The action surprise model explains several features of midbrain dopamine activity and also makes several testable predictions, which we outline below.

\emph{\textbf{Nonspecific encoding of action}}. A central property of the model is that the action surprise term $\norm{\vec{a}_t - \vec{\mu}(\vec{s}_t)}^2$ is not action-specific---it reflects only scalar information about the agent's action, even in high-dimensional action spaces, and does not distinguish between two equally surprising actions. This contrasts with the representation in striatal projection neurons forming the output of the actor network, which specify movement commands.  Indeed, experimental recordings have found detailed encoding of kinematics in striatal projection neurons \citep{dhawale2021basal} but only coarse movement-related signals in dopamine activity that do not reliably distinguish between movement types \citep{da2018dopamine, mendoncca2021transient}. This distinction is consistent with our model and inconsistent with models that explain action-modulated dopamine activity in terms of specific motor commands.

\emph{\textbf{Decrease in action-modulated dopamine activity with learning}}. The action surprise model predicts that movement-related dopamine activity is lower when the basal ganglia's policy more closely matches the agent's actions.  Before the basal ganglia has learned an effective policy, actions driven by an expert external controller will typically provoke large action surprise.  Once the basal ganglia network has learned an expert policy itself, action surprise will typically be lower.  Thus, we predict a reduction in the magnitude of action surprise signals over the course of learning. We observed this phenomenon in both simulated tasks (Fig.~\ref{fig:predictions}A).  This prediction is supported by a finding that dopamine activity coinciding with lever-press movements decreases with repeated task practice \citep{da2018dopamine}.

\emph{\textbf{Context-dependent correlation between dopamine activity and acceleration}}.  Several studies have shown that striatal dopamine activity correlates with movement speed and acceleration \citep{wang2011conjunctive, puryear2010conjunctive, barter2015beyond}. Others \citep{howe2016rapid, da2018dopamine, mendoncca2021transient} have shown that activity in movement-responsive midbrain dopamine neurons is modulated primarily movement at initiation (and in some cases at movement offset) but less so during ongoing movements.  Our model gives rise to a complex relationship between movement and action-driven dopamine signals, with the two appearing correlated on some trials but not others (Fig.~\ref{fig:predictions}B).  On average, action-driven dopamine activity is highest at trial onset (Fig.~\ref{fig:predictions}C), and it is positively correlated with the agent's acceleration (Fig.~\ref{fig:predictions}D).  Our results suggest that the representations of movement initiation and kinematic information attributed to dopamine activity in prior work may actually emerge as byproducts of an action surprise computation.

\emph{\textbf{Neural circuits underlying the computation of action surprise}}. The action surprise model requires information about the agent's action be available to midbrain dopamine neurons for computation of the action surprise term.  This is consistent with the presence of pathways from motor cortex and cerebellum to midbrain dopamine neurons \citep{watabe2012whole, menegas2015dopamine}. We note that the specific form of the action surprise term in our model, $\norm{\vec{a} - \vec{\mu}(\vec{s})}^2$, is somewhat arbitrary, and other measures of distance between $\vec{a}$ and $\vec{\mu}(\vec{s})$ may be equally suitable.  We also note that our model is agnostic as to whether the action surprise and RPE components of dopamine activity are represented in the same individual neurons.  Empirically, RPE-signalling and movement-signaling dopamine neurons appear to comprise distinct but partially overlapping populations in the same striatal subregions \citep{da2018dopamine}, though data on this question is limited.

\section{Discussion}

Our work demonstrates a new link between action-related midbrain dopamine activity and reinforcement learning in the basal ganglia. While each of these topics has received extensive treatment in the neuroscience literature, they have typically been studied separately.  Midbrain dopamine activity is known to causally affect movement initiation and invigoration in real time \citep{da2018dopamine, panigrahi2015dopamine}.  Consequently, the function of such activity is often regarded as a motivational signal \citep{wise2004dopamine, mohebi2019dissociable} separate from dopamine's role in reinforcement learning. However, action and RPE-related dopamine release coincide in the same striatal subregions, often even in the same neurons, and with comparable magnitude \citep{wang2011conjunctive, puryear2010conjunctive, da2018dopamine, mendoncca2021transient, engelhard2019specialized}. Thus, it is likely that both action-related and RPE-related dopamine activity impact cortico-striatal synaptic plasticity and learning. Our results show that such an influence is a necessary component of an architecture capable of off-policy learning.

The effects of action-related dopamine activity, which is not a feature of classic RL models of the basal ganglia, may be more easily overlooked in the context of simple tasks in which animals learn to associate a small set of cues or actions with immediate reward. In such tasks, standard RPE-based models are adequate to achieve good performance.  The failure of such models that we observe, which arises from a mismatch between the value functions associated with the behavioral and basal ganglia policies, only emerges in tasks where reward is obtained through an extended sequence of actions.

Our model suggests that distributed control of behavior by many regions plays an important role in learning in the basal ganglia. A number of experiments have observed differential contributions from multiple brain regions during learning. For instance, some motor skills are observed to recruit the motor cortex early in learning before being consolidated into the basal ganglia \citep{kawai2015motor, hwang2019disengagement}.  These results suggest that regions other than the basal ganglia may be more adept at flexibly adapting to novel tasks, while the basal ganglia specializes in consolidating well-practiced skills.  Other studies have observed parallel contributions of model-based and model-free reinforcement learning strategies in the same task \citep{smittenaar2013disruption, glascher2010states}, revealing arbitration mechanisms that transfer control between them in order to leverage the advantages of each \citep{lee2014neural, kim2019task}.  The flexibility afforded by off-policy RL algorithms, such as the action surprise model, enables the basal ganglia to benefit from complementary learning and control strategies adopted by other neural circuits.  Exploring how off-policy RL algorithms can best leverage these diverse sources of expertise is a fruitful avenue for extensions to our model.  For instance, action surprise signals may be modulated by the confidence of external controllers in order to more efficiently learn from expert behavior. 

Predictions of the action surprise model can be tested by experimental recordings of striatal projection neurons, midbrain dopamine neurons, or other motor areas.  Our work also has implications for reinforcement learning.  The introduction of a biologically plausible off-policy reinforcement learning algorithm, involving local learning rules and a single modulatory factor, enables the deployment of off-policy RL on plasticity-enabled neuromorphic hardware \citep{davies2018loihi}.  Moreover, we anticipate that insights into how neural circuits learn from off-policy behavior governed by distributed and diverse controllers will provide useful inspiration for RL algorithms.

\begin{ack}

This work was supported by NSF NeuroNex Award DBI--1707398 and The Gatsby Foundation (GAT3708).  ALK was also supported by the McKnight, Burroughs-Wellcome, and Mathers Foundations.  JL was also supported by the DOE CSGF (DE--SC0020347).  The authors declare no competing interests.

\end{ack}

\bibliographystyle{plain}
\bibliography{bibliography}
\newpage
\appendix

\section{Appendix}

\subsection{Reinforcement learning notation}
\label{appendix:notation} 
We adopt standard reinforcement learning notation \citep{sutton2018reinforcement}.  An agent progresses through a sequence of states $\vec{s}_t$ by executing actions $\vec{a}_t$ which influence transition probabilities between states.  We focus on the case where actions are continuous, vector valued quantities $\vec{a}_t \in \mathbb{R}^D$, where $D$ is the dimension of the action space.  The agent's policy---the probability distribution from which it samples actions given the state $\vec{s}$---is written as $\pi(\vec{a} | \vec{s})$.  States are associated with scalar rewards $r_t = r(\vec{s}_t)$.  The \emph{value function} associated with a policy $\pi$ is defined as follows

\begin{align}
V_\pi(\vec{s}) = E_{\pi}\left[\sum_{t=0}^{T} \gamma^t r_t\right].
\end{align}

Where the expectation is taken over the policy $\pi$ and transitions of the environment, and conditioned on $\vec{s}_{t=0} = \vec{s}$, $\gamma \in [0, 1]$ is a discount factor, and $T$ indicates the potentially finite horizon of an episode.  An agent’s goal is to learn a policy $\pi$ that maximizes $V_\pi(\vec{s})$.  We are often also interested in the \emph{action-value} function $Q_\pi(\vec{s}, \vec{a})$, defined exactly the same as $V_\pi(\vec{s})$ except the sum is also conditioned on the initial action $\vec{a}_{t=0} = \vec{a}$.  The $Q$-learning algorithm in Eq.~\ref{eq:Qlearning} learns an estimate $\hat{Q}(\vec{s}, \vec{a})$ of the $Q$-function under the assumption that the current policy $\pi$ to be deterministic such that the action selected in state $\vec{a}$ is always $\mathrm{argmax}_{\vec{a}'} \hat{Q}(\vec{s}, \vec{a}')$.

\subsection{Biological implementations of three-factor learning rules for continuous action spaces}
\label{appendix:threefactor} 

The three-factor learning rule in Eq. ~\ref{eq:threefactor} and the analogous rule for the action surprise model involve a product of a presynaptic term $\phi_j(\vec{s}_t)$, a dopaminergic term $\delta_t$ (or $\delta_t^+$ in the action surprise model model), and a postsynaptic factor $(\vec{a}_t)_i - \mu_i(\vec{s}_t)$.  Three-factor plasticity rules involving pre-synaptic, post-synaptic, and neuromodulatory factors have been observed experimentally \citep{ wickens1996dopamine, calabresi2000dopamine, contreras1999predictive} and are commonly used in RL modeling \citep{sutton2018reinforcement, niv2009reinforcement}.  However, the difference  $(\vec{a}_t)_i - \mu_i(\vec{s}_t)$ involved in the postsynaptic term, which arises in our framework from the use of a continuous action space, requires more biological justification. Note that $\mu_i(\vec{s}_t)$ is the contribution to to the postsynaptic neuron's activity driven by the cortico-striatal synapses subject to the RL algorithm, while $(\vec{a}_t)_i$ takes into account exploration noise and influence from external controllers.  Biologically, the difference between these terms may contribute to the plasticity in a number of ways, which we summarize here.

\textbf{Option 1: Multi-compartment neurons}. Potentially, the original basal ganglia-driven action signal $\mu_i(\vec{s}_t)$ and the difference signal $(\vec{a}_t)_i - \mu_i(\vec{s}_t)$ could arrive at different dendritic compartments.  In this case, the total activity of the neuron would reflect the efferent copy $(\vec{a_i})_t$, but a compartment-specific plasticity rule would enable the synaptic weight update to depend only on the appropriate term.

\textbf{Option 2: Time-varying striatal activity and a temporal plasticity kernel}. If striatal projection neuron activity initially represents $\mu_i(\vec{s}_t)$ before receiving additional inputs which cause it to represent $(\vec{a}_t)_i$, then a spike-timing dependent three-factor learning rule with a suitable temporal kernel can result in an update that makes use of the difference in activity between the two phases.  Indeed, such temporal kernels have been observed at cortico-striatal synapses \citep{shen2008dichotomous}.  

\textbf{Option 3: Normalization}.  Alternatively, the plasticity rule may only explicitly depend on projection neuron activity representing $(\vec{a_t})_i$, with normalization mechanisms across the striatal population implicitly contributing the $-\mu_i(\vec{s}_t)$ term.  For concreteness, suppose the $i$th action dimension of the basal ganglia network policy $\mu_i(\vec{s})$ is a linear function of the \emph{normalized, nonnegative} firing rates $\vec{x}$ of the striatal projection neuron population in response to $\vec{s}$:

\begin{align}
\mu_i(\vec{s}) = \frac{\vec{w} \cdot \vec{x}(\vec{s})}{\sum_i{x_i(\vec{s})}}.
\end{align}

Further suppose that as the agent takes an action, the population activity is updated to $\vec{y}$ to reflect an efferent copy of the action $\vec{a}$:

\begin{align}
a_i = \frac{\vec{w} \cdot \vec{y}(\vec{s})}{\sum_i{y_i(\vec{s})}}.
\end{align}

If cortico-striatal synapses are updated according to a three factor learning rule $\Delta \vec{w} \propto \delta \vec{\phi}(\vec{s}) \vec{y}(\vec{s})$, then at subsequent occurrences of state $\vec{s}$, the basal ganglia network will drive will evoke population activity:

\begin{align}
\vec{x}^{\mathrm{new}} = \vec{x} + \epsilon \delta \vec{y},
\end{align}

for some learning rate $\epsilon$.  Note that, taking into account the normalization mechanism, the directional derivative of $\mu_i$ with respect to $\vec{x}$ along $\vec{y}$ is

\begin{align}
D_{\vec{y}} \mu_i(\vec{x}) = \frac{\left(\sum_j x_j(\vec{s})\right) \vec{w} - (\vec{w} \cdot \vec{x})\vec{1}}{\left(\sum_j x_j(\vec{s})\right)^2}.
\end{align}

And so following the weight update the action $\mu_i^{\mathrm{new}}$ driven by the basal ganglia network in state $\vec{s}$ will be

\begin{align}
&(\mu_i)^{\mathrm{new}}(\vec{s})\\
&= (\mu_i)^{\mathrm{old}}(\vec{s}) + \epsilon \frac{(\sum_i x_i(\vec{s})) \vec{w} \cdot \vec{y} - (\vec{w} \cdot \vec{x})\vec{1} \cdot \vec{y}}{(\sum_i x_i(\vec{s}))^2} \\
&= (\mu_i)^{\mathrm{old}}(\vec{s}) + \epsilon \delta \frac{ \vec{w} \cdot \vec{y} - (\vec{w} \cdot \vec{x})\vec{1} \cdot \vec{y}}{(\sum_i x_i(\vec{s}))} \\
&= (\mu_i)^{\mathrm{old}}(\vec{s}) + \epsilon \delta \frac{{(\sum_i y_i(\vec{s}))}}{{(\sum_i x_i(\vec{s}))}} \left(\frac{\vec{w} \cdot \vec{y}}{\sum_i y_i(\vec{s})} - \frac{\vec{w} \cdot \vec{x}}{\sum_i x_i(\vec{s})}\right) \\
&= (\mu_i)^{\mathrm{old}}(\vec{s}) + \epsilon \delta \frac{{(\sum_i y_i(\vec{s}))}}{{(\sum_i x_i(\vec{s}))}} \left(a - \mu(\vec{s})\right)\\
&= (\mu_i)^{\mathrm{old}}(\vec{s}) + c \delta \left(a - \mu(\vec{s})\right).
\end{align}
for some constant $c$, which up to a scalar factor is exactly the update that would be induced by Eq. ~\ref{eq:threefactor} in the absence of a normalization mechanism.  Note that the update to the cortico-striatal weights does not necessarily follow the policy gradient in weight space, but we have shown that it induces the appropriate update to the basal ganglia network policy.

\subsection{Alternative approaches to off-policy learning in continuous action spaces}
\label{appendix:offpolicy} 

A popular approach to continuous off-policy deep RL, adopted in the DDPG \citep{lillicrap2015continuous} and SAC \citep{haarnoja2018soft} algorithms, is to parameterize $\hat{Q}(\vec{s}, \vec{a})$ with a neural network taking $\vec{s}$ and $\vec{a}$ as inputs, and to use a second neural network to learn $\mathrm{argmax}_{\vec{a}'} \hat{Q}(\vec{s}, \vec{a}')$ (or, in the case of SAC, a probability distribution peaked at this value).  This second network can be used both as the policy network and for the purpose of computing  the factor $\max_{\vec{a}'} \hat{Q}(\vec{s}, \vec{a}')$ used in updates to $\hat{Q}$.  While this strategy has proven to be effective in deep RL, a biological implementation is not readily apparent.  In particular, in this approach, the parameter updates for the actor require computing $\nabla_{\vec{a}} Q(\vec{s}, \vec{a})$, which depends a substantial amount of information that is not local to the actor.  By contrast, in our approach, updates to the actor network depend on the critic network only through the scalar factor $\delta$, which biologically can be signalled by dopamine release. Previous studies have suggested that a quadratic approximation to the $Q$-function, as we implement in our model, can approach or exceed the performance of DDPG \cite{gu2016continuous}.

\subsection{Learning confidence parameters}
\label{appendix:confidence} 

\begin{figure}[h!]
    \label{fig:learn_sigma}
    \centering
    \includegraphics{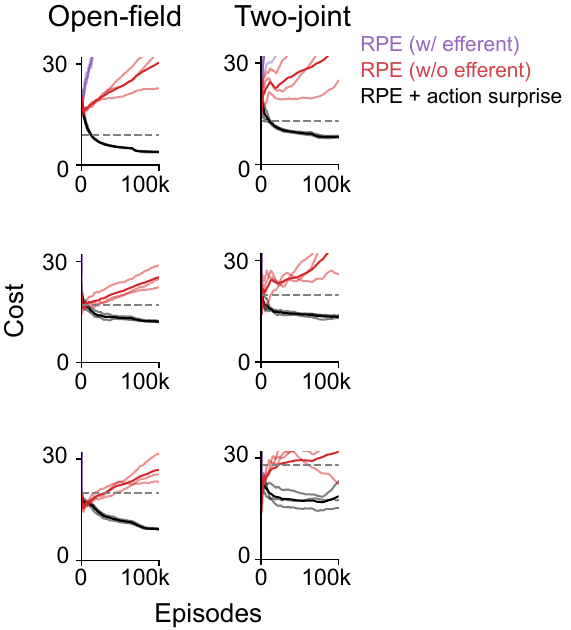}
    \caption{Same information as Fig.~\ref{fig:results2}C, but in this case the black trace was learned with the action surprise coefficient $\frac{1}{\sigma^2}$ learned online, following the update rule Eq. \ref{eq:sigma_update}.}
    \label{fig:averaging}
\end{figure}

In the on-policy case introduced in Section \ref{section:background}, following the policy gradient (Eq.~\ref{eq:policygradient}) yields the following update for the variance parameter $\sigma$:

\begin{align}\Delta \sigma \propto -\frac{1}{\sigma^2} + \frac{2}{\sigma^3}  \delta_t\norm{\vec{a}_t - \vec{\mu}(\vec{s}_t)}^2.
\end{align}

In the action surprise model,  under the parameterization of Eq.~\ref{eq:parameterization}, following gradient updates of the loss function of Eq.~\ref{eq:Qlearning} gives the following update for the action surprise coefficient $\sigma$:

\begin{equation}
\label{eq:sigma_update}
\Delta {\sigma}= \frac{2}{\sigma^3}\delta_t^+ \norm{\vec{a}_t - \vec{\mu}(\vec{s}_t)}^2.
\end{equation}

Note that this update is exactly the same as in the off-policy case but without the decay term $-\frac{1}{\sigma^2}$ providing an impulse toward increasing confidence (decaying $\sigma$) with increasing training data.  This difference reflects the fact that the role of $\sigma$ in the $Q$-learning model algorithm is not to measure uncertainty as in the on-policy case, but rather to approximate the width of the $Q$-function, which will have some finite value independent of the amount of data observed.

In Fig ~\ref{fig:learn_sigma} we show the result of using this update rule to learn the coefficient $\sigma$ online (after initializing $\frac{1}{\sigma^2}$ at 0, rather than optimizing it as a hyperparameter) for the full off-policy learning case.  The model is capable of converging to an appropriate $\sigma$ such that it learns successfully, while (as discussed in the main text) the RPE-only models are incapable of learning.

\subsection{Further simulation details}
\label{appendix:parameters} 

For all experiments, we optimized hyperparameters over the following ranges: exploration noise variance $\in \{0.5, 1.0, 2.0, 4.0, 8.0\}$, learning rate $\in \{0.05, 0.1, 0.2\}$.  The critic learning rate was always set to $0.1$.  For the RPE-only algorithms, the  learning rate of the actor was optimized over $\{0.0315, 0.0625, 0.125, 0.25, 0.5 \}$, and for the action surprise model, the value of $\frac{1}{\sigma^2}$ was optimized over the same set.  The velocity and acceleration penalties in the task cost functions, $\alpha_1$ and $\alpha_2$, were each always set to $0.1$, and the discount factor $\gamma$ always set to 0.99.

For all simulations, training was conducted for 100000 episodes, each with a length of 10 timesteps.  Each training run was performed on a single NVIDIA GeForce RTX 2080 Ti GPU on an internal cluster.

\subsection{Results for averaging-based combination of basal ganglia network and external controller policies}
\label{appendix:averagingresults} 

\begin{figure}[h!]
    \centering
    \includegraphics{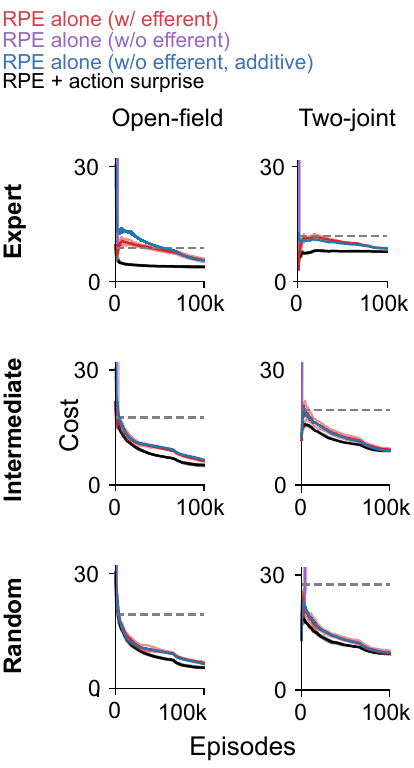}
    \caption{Same information as Fig.~\ref{fig:results2}B, for the case where the basal ganglia policy and external controller policy are combined by taking their mean, rather than sampling.  Additionally blue trace shows an alternative in which output of the basal ganglia network is added to (rather than averaged with) the external controller policy.}
    \label{fig:averaging}
\end{figure}

\end{document}